\documentclass[sigconf]{acmart}

\setcopyright{none}
\renewcommand\footnotetextcopyrightpermission[1]{} % removes footnote with conference information in first column
\AtBeginDocument{%
  \providecommand\BibTeX{{%
    \normalfont B\kern-0.5em{\scshape i\kern-0.25em b}\kern-0.8em\TeX}}}

\usepackage{booktabs}
\usepackage{mathrsfs}
\usepackage{multirow}
\usepackage{subfig}
\usepackage{bm}
\usepackage{float}
\usepackage{color}
\usepackage{makecell}
\usepackage{listings}
\begin{document}

\title{
Efficient Sparse Matrix Kernels based on Adaptive Workload-Balancing and Parallel-Reduction 
}
\author{Guyue Huang$^{1*}$, Guohao Dai$^2$, Yu Wang$^2$, Yufei Ding$^1$ and Yuan Xie$^1$}
\affiliation{$^1$University of California, Santa Barbara $^2$Tsinghua University}
\email{*guyue@ucsb.edu}

\begin{abstract}
  Sparse matrix-vector and matrix-matrix multiplication (SpMV and SpMM) are fundamental in both conventional (graph analytics, scientific computing) and emerging (sparse DNN, GNN) domains. Workload-balancing and parallel-reduction are widely-used design principles for efficient SpMV. However, prior work fails to resolve how to \textit{implement} and \textit{adaptively use} the two principles for SpMV/MM. To overcome this obstacle, we first complete the implementation space with optimizations by filling three missing pieces in prior work, including: (1) We show that workload-balancing and parallel-reduction can be combined through a segment-reduction algorithm implemented with SIMD-shuffle primitives. (2) We show that parallel-reduction can be implemented in SpMM through loading the dense-matrix rows with vector memory operations. (3) We show that vectorized loading of sparse rows, being a part of the benefit of parallel-reduction, can co-exist with sequential-reduction in SpMM through temporally caching sparse-matrix elements in the shared memory. In terms of adaptive use, we analyze how the benefit of two principles change with two characteristics from the input data space: the diverse sparsity pattern and dense-matrix width. We find the benefit of the two principles fades along with the increased total workload, i.e. the increased dense-matrix width. We also identify, for SpMV and SpMM, different sparse-matrix features that impact workload-balancing effectiveness. Our design consistently exceeds cuSPARSE by 1.07-1.57$\times$ on different GPUs and dense matrix width, and the kernel selection rules involve 5-12\% performance loss compared with optimal choices. Our kernel is being integrated into popular graph learning frameworks~\cite{dgl,cogdl} to accelerate GNN training. \footnote{This project is available at https://github.com/hgyhungry/dgSPARSE-Library}
\end{abstract}

\maketitle

\section{Problem and Motivation} \label{sec:intro}

\begin{figure}[t]
  \includegraphics[width=\linewidth]{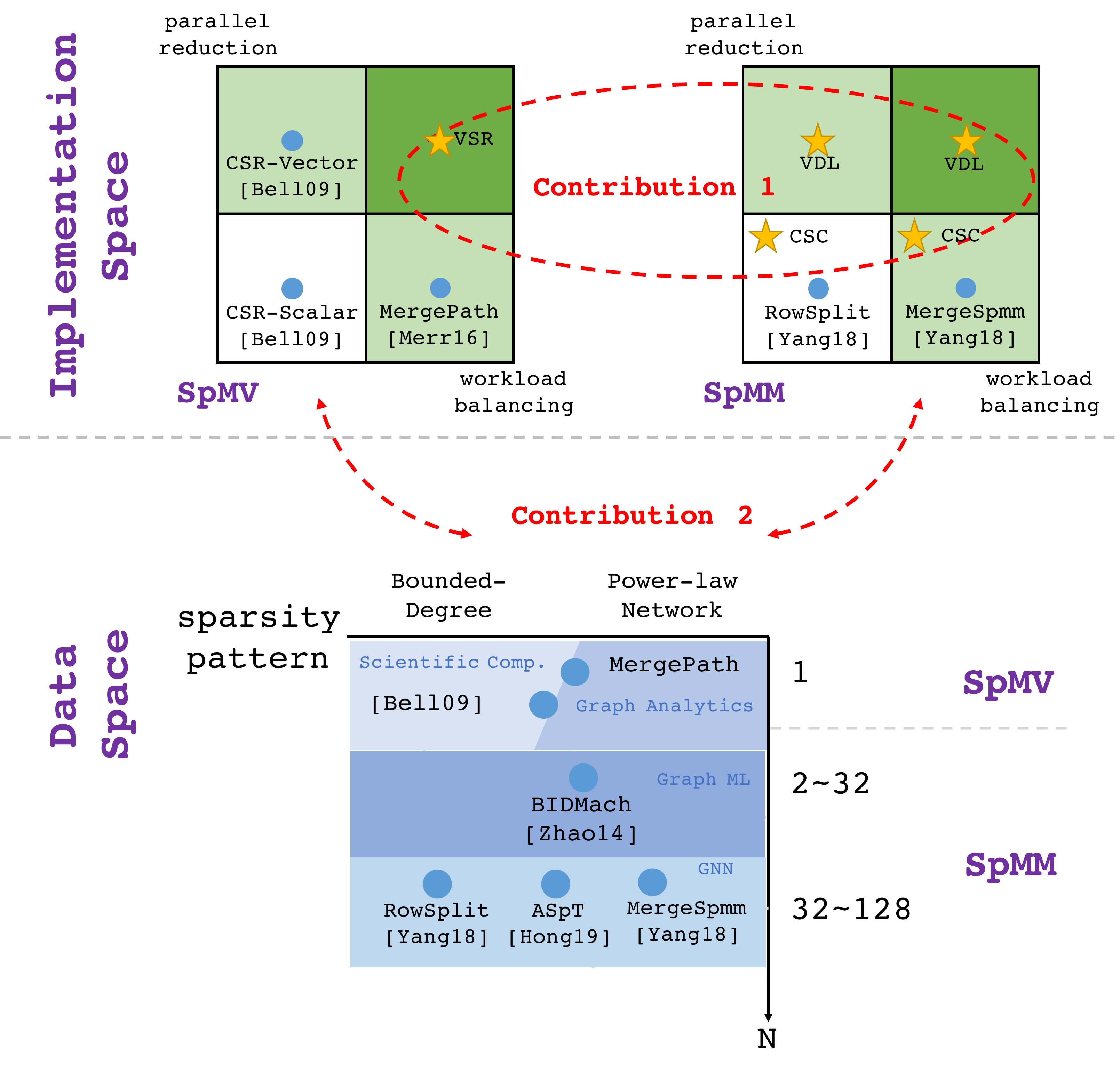}
  \caption{Target problem and contributions of this work. We consider two design principles, workload-balancing and parallel-reduction. First, we complete the implementation space with highly-optimized kernels. Second, we map a variety of SpMV/MM problems to implementations adaptively. }
  \label{fig:motivation}
\end{figure}

Efficient basic sparse-matrix primitives can benefit a variety of applications. The sparse matrix multiplication $Y_{M \times N} = A_{M \times K}X_{K \times N}$ where $A$ is sparse and $X,Y$ are dense, is referred to as Sparse Matrix-Vector product (SpMV, when $N=1$) or Sparse Matrix-Matrix product (SpMM, when $N>1$). SpMV and SpMM are fundamental components to a wide range of problem domains. SpMV is used in graph analytics and scientific computing~\cite{ yang2011fast, 10.5555/829576}. SpMM is used in iterative algorithms for sparse matrix factorization~\cite{10.5555/3154630.3154684}. Recent advances in sparse NN, promising higher computational efficiency than dense models, rely on fast SpMV/MM kernels to demonstrate speedup in practice~\cite{sputnik}. SpMM is also a core operation in graph neural networks (GNNs)~\cite{dgl2,hu2020featgraph}. Accelerating SpMV/MM on GPUs, the dominating HPC hardware in presence, can potentially boost the performance of many aforementioned applications. 

\begin{figure}[t]
  \includegraphics[width=\linewidth]{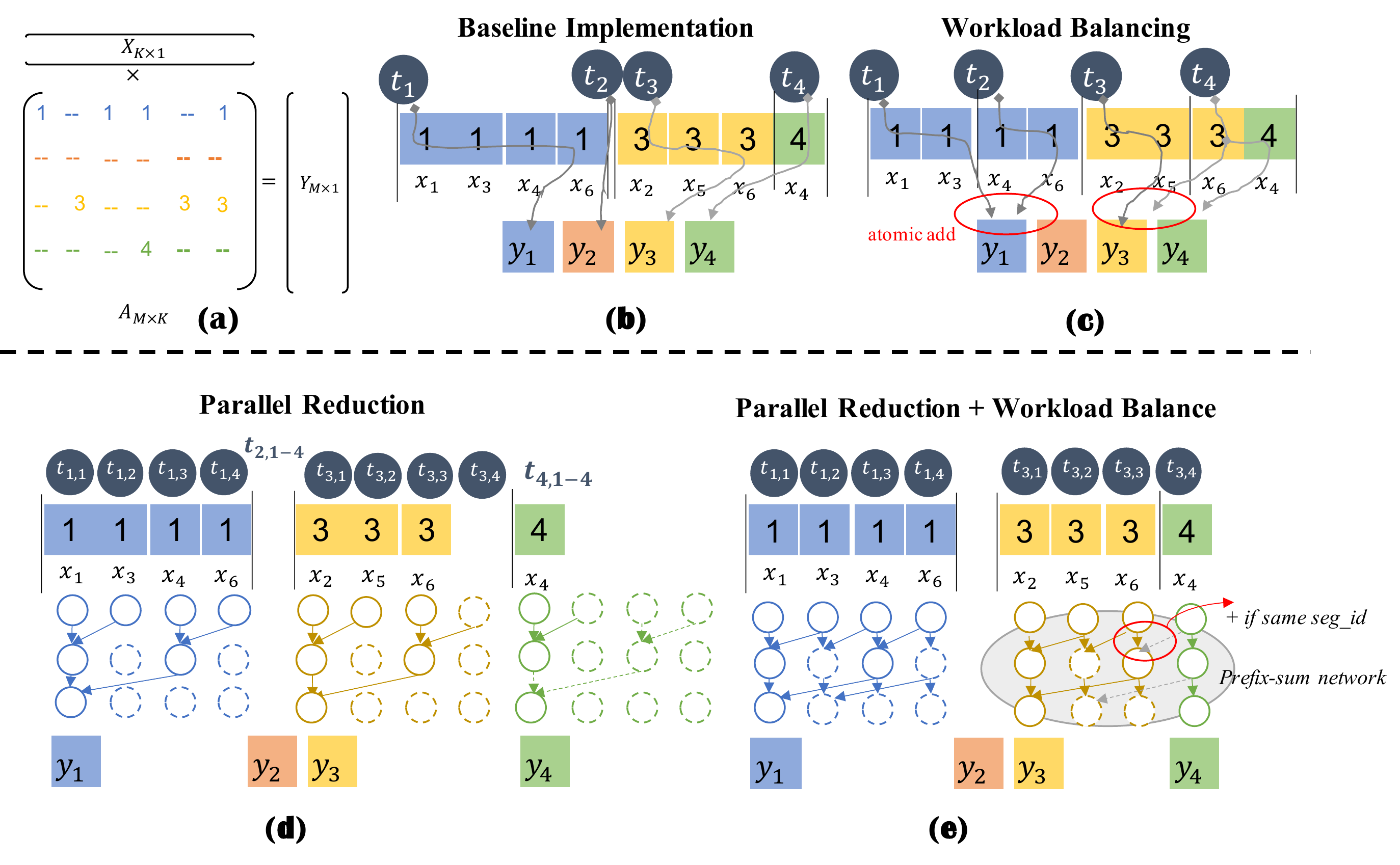}
  \caption{Illustration of workload-balancing and parallel-reduction. We combine the two principles with a vectorized segment-reduction algorithm with SIMD-shuffle primitives as shown in (e). }
  \label{fig:spmv}
\end{figure}

\textbf{Workload-balancing} and \textbf{parallel-reduction} are two design principles extensively studied for SpMV. Workload-balancing prevents the kernel from being bottlenecked by  the most work-intensive thread, either through row-binning~\cite{greathouse2014efficient,sputnik}, or through fine-grained segmentation of arithmetic and memory operations~\cite{merrill2016merge}(Figure.~\ref{fig:spmv}(b)). Parallel-reduction increases the resource occupancy and bandwidth utilization compared to a sequential-reduction, through performing inner-product with vectorized element-wise multiplication and SIMD merge-tree~\cite{bell2009implementing}(Figure.~\ref{fig:spmv}(c)). 
% Workload-balancing and parallel-reduction are used in vendor library~\cite{cusparse} for SpMV. 
Prior art, however, leaves some missing pieces in the implementation and adaptive use of the two principles.

In terms of implementation, we lack the guidance to combine the two principles and extend them from SpMV to SpMM. Firstly, the combination of workload-balancing and parallel-reduction is not covered by prior work. The combination leads to a segment-reduction operation, whose existing solution is through tiled sequential scan, with limited resource occupancy. Secondly, how to apply parallel-reduction in SpMM is missed in previous work. Previous work on SpMM~\cite{graphblas, aspt} all apply sequential-reduction, and target $N\geq 32$ scenarios. Given the effectiveness of parallel-reduction in SpMV~\cite{bell2009implementing}, we naturally ask if parallel-reduction can bring similar benefit to SpMM with small $N$. However, prior work does not cover how to perform parallel-reduction in SpMM. We propose efficient implementations for these missing pieces, as marked in Figure.~\ref{fig:motivation}.

In terms of adaptive use of workload-balancing and parallel-reduction based on input characteristics, prior work fails to solve two problems: firstly, how to selectively apply the two principles based on the \textbf{sparsity pattern}, and secondly, how the answer to the first question changes with the \textbf{dense-matrix width $N$}. Many prior researches apply work-balancing not always but selectively. Still, comprehensive analysis of how sparsity features and $N$ affect this choice is missing in prior work. Parallel-reduction is very effective for SpMV~\cite{bell2009implementing}, but not applied to SpMM designs~\cite{graphblas,aspt,gespmm,sputnik}. The transition point from parallel-reduction to sequential-reduction is also not discussed in previous work. 
% To illustrate the first issue, we implement the two principles on top of a simple SpMV algorithm CSR-Scalar~\cite{bell2009implementing}. Observe from Figure.~\ref{fig:motivation} that workload balancing and parallel reduction bring up to \textbf{73.23$\times$,19.43$\times$} speedup for some matrices, but also up to \textbf{10.35$\times$,9.09$\times$} slowdown on others. 
%  The second issue comes with SpMM scenarios, when we must re-consider how to implement and how to adaptively use workload-balancing and parallel-reduction. 

Rethinking the two design principles for SpMV/MM, we make the following contributions:
\begin{itemize}
  \item \textbf{Complete space with optimizations}. We fill the gaps of practicing workload-balancing and parallel-reduction in SpMV/MM with three novel optimizations: vectorized segment reduction, vector-type dense-row loading, and coalesced sparse-row caching. (Section.~\ref{sec:approach:impl})
  \item \textbf{Heuristics from data to implementation}. We draw insights from evaluating the two principles on a large benchmark of sparse matrices, and to a range of $N$ from $1$ up to $128$. We analyze why the benefit of two principles fades as $N$ increases, and provide low-cost rules to selectively apply them. (Section.~\ref{sec:approach:adaptive})
  \item \textbf{Comprehensive experimental results}. We extensively evaluate our method on three GPUs. Our approach outperforms cuSPARSE~\cite{cusparse} by\textbf{ 1.07-1.52$\times$}. The kernel selection strategy demonstrates an average 5-12\% performance loss, compared to 68\% in minimum if always picking one design. (Section.~\ref{sec:results})
\end{itemize}

\section{Approach and Uniqueness} \label{sec:approach}

\subsection{Implementing Two Principles}\label{sec:approach:impl}

In this part, we present three optimizations for best practice of workload-balancing and parallel-reduction in SpMV/MM.

\subsubsection{Vectorized Segment Reduction (VSR)}

\textbf{Motivation: } The combination of workload-balancing parallel-reduction can effectively handle \textit{short rows} in the sparse matrix. As shown in Figure.~\ref{fig:spmv}(d), in CSR-Vector SpMV, the de-facto practice of parallel-reduction, when the number of non-zeros in a row is smaller than the number of threads in a GPU warp (similar to a SIMD thread bundle), parallel-reduction performs many wasted operations. On the contrary, warps processing long rows bear heavy workload and can bottleneck the execution. To solve this, we apply the workload-balancing principle by assigning to each warp a fixed number of non-zeros instead of a certain row, as in Figure.~\ref{fig:spmv}(e). 

\textbf{Approach: } After applying workload-balancing, the elements assigned to each warp can cross the boundary of rows. Hence, instead of the pure merge-tree reduction in CSR-Vector, we need to perform a segment reduction operation referring to the row-indices of each element. We implement our vectorized segment reduction (VSR) algorithm by simulating a prefix-sum network, but the reduction operation is to \textit{add if the indices of two elements match}. Finally, we make each thread compare their indices with their right-side neighbor, to detect if they are the start of a segment and must dump out their results.

\textbf{Results: } \footnote{All ablation study in this section is conducted on RTX3090. We present overall results for three GPUs in Section.~\ref{sec:results}.} We compare the VSR design with three baselines: SpMV without workload-balancing or parallel-reduction, and with each single optimization. When tested on the SuiteSparse~\cite{10.1145/2049662.2049663} benchmark, VSR-based SpMV exceeds the other three kernels (baseline and two optimizations individually) on 40.8\% of the matrices.  

\subsubsection{Vector-type Dense-row Loading (VDL)}

\textbf{Motivation: } This is an optimization on the straightforward implementation of parallel-reduction SpMM. Based on parallel-reduction SpMV, parallel-reduction SpMM can be completed with $N$-times SpMV on every column of the dense matrix. However, this straightforward implementation suffers from inefficient memory operations. In parallel-reduction SpMV, threads in a warp load dense-vector elements according to positions of the non-zeros in the sparse row. The addresses of target dense elements are not contiguous, leading to poor locality. 
%Consider SpMM, since in most applications the dense matrix is stored in a row-major fashion, the target addresses are further separated by the elements in the same row like in Figure.~\ref{}. Threads will load from addresses that spread over a large space, which limits the bandwidth efficiency. 

\begin{figure}[t]
    \includegraphics[width=0.7\linewidth]{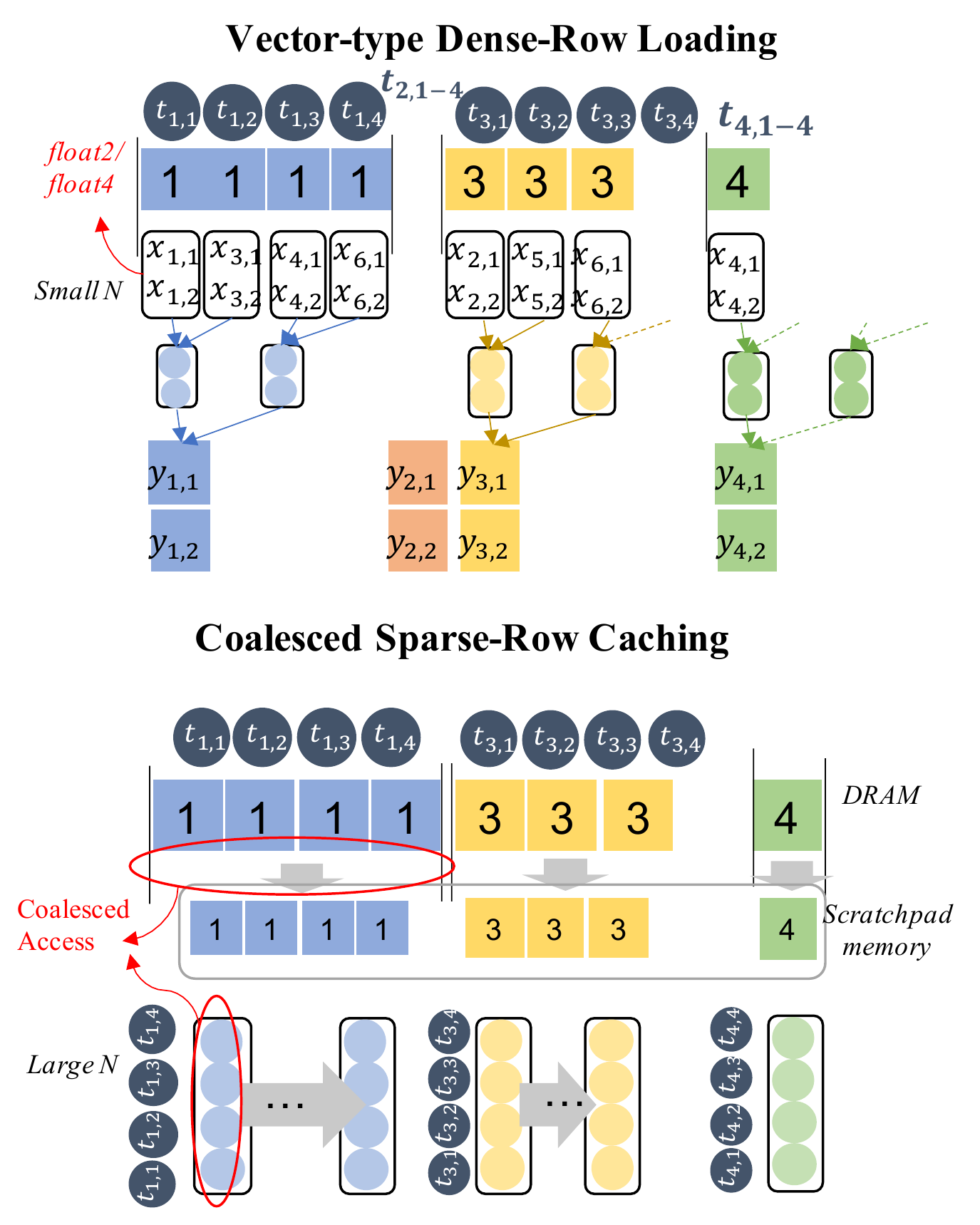}
    \caption{Illustration of VDL and CSC. 
    % VDL exploits vector types like float2/float4 to increase the amount of effective data per request. CSC enables coalesced loading of both sparse and dense elements through buffering sparse elements in scratchpad memory.
    }
    \label{fig:spmm}
\end{figure}

\textbf{Approach: } We explore the insight that each sparse-matrix element $A[i, k]$ can be multiplied with all elements in a dense-matrix row, i.e. $X[k, n]$ for $1\geq n \leq N$. Our design is shown in Figure.~\ref{fig:spmm}. We make each thread load multiple dense-matrix elements using the vector-type memory operations, i.e. loading with type \textit{float2/float4}. Take \textit{float2} as an example, each thread holding a sparse non-zero $A[i,k]$ will load $X[k, 0], X[k,1]$ together in one instruction, multiply $A[i,k]$ with both, and accumulate the results on two different partial-sums. This design can increase the amount of effective data per request, at least the size of \textit{float2/float4}.

\textbf{Results: } We set $N=2$ and compare \textit{float2}-loading VDL against performing two SpMVs. We synthesize 27 matrices with the R-MAT generator~\cite{chakrabarti2004r} using various size, sparsity and distribution parameters. On this micro benchmark, VDL performs 1.89$\times$ better than the two-SpMV solution. 

\subsubsection{Coalesced Sparse-row Caching (CSC)}

\textbf{Motivation: } Observe there are two benefits of parallel-reduction against sequential-reduction. Firstly, loading the sparse elements is more efficient. Threads in a warp load non-zeros in a sparse row that are stored contiguously, which is an ideal data access pattern on GPUs. Secondly, arithmetic operations are performed by more threads to exploit parallel resources. SpMM has plenty of parallelism, making the parallelized arithmetic operation less necessary. However, we would like to keep the benefit of efficient load operations.

\textbf{Approach: } We implement vectorized loading of sparse rows under a sequential-reduction scheme by exploiting the shared memory. The shared memory is a scratchpad memory accessible by all threads within the same warp. We first load the non-zeros in one row in a coalesced way, i.e. load $warp\_size$ non-zeros with one instruction. We store these elements into the shared memory. Next, we make parallel threads iterate over the cached elements, and compute on different columns of the dense matrix, as in Figure.~\ref{fig:spmm}.

\textbf{Results: } We set $N=128$ and compare CSC against SpMM implementation with pure sequential-reduction. On the micro benchmark described in the previous part, CSC brings average 1.20$\times$ speedup. 

\subsection{Adaptive Two Principles To Problems} \label{sec:approach:adaptive}

The previous part introduces our novel methods to optimize workload-balancing and parallel-reduction for SpMV and SpMM. This part introduces our second contribution, which is a kernel selection strategy for various inputs, i.e. the sparsity pattern and $N$. We derive our selection algorithm from the following three insights:

\begin{itemize}
    \item \textbf{Insight 1}: Parallel-reduction is necessary for efficiently loading the sparse matrix, but suffers from poor memory locality in loading the dense matrix when $N$ gets larger.
    \item \textbf{Insight 2}: Workload-balancing is necessary for matrices with skewed non-zero distribution, but involves extra overhead for well-balanced matrices.
    \item \textbf{Insight 3}: A common benefit of two principles, the increased parallelism and resource occupancy, becomes unnecessary when the total amount of work is large, either because of large size of the sparse matrix, or because of large $N$.
\end{itemize}

\textbf{Insight 1} states how $N$ affects whether to use parallel-reduction. As discussed in our VDL optimization in the previous part, parallel-reduction benefits from efficient loading of the sparse matrix, while exploiting locality in the dense matrix when non-zeros are clustered. However, when $N$ increases, parallel-reduction suffers from poor locality when loading elements from the dense matrix if $N$ is large. On the contrary, sequential reduction benefits from a friendly access pattern to the dense matrix. Enhanced with our CRC optimization, sequential-reduction outperforms parallel-reduction by a large margin due to efficient memory operations.

\textbf{Insight 2} states how the sparsity pattern affects whether to use workload-balancing. Intuitively, if a sparse matrix has severely imbalanced non-zero distribution, parallelizing different rows to different threads results in imbalanced workloads, which is why workload-balancing is necessary. Statistical metrics such as standard deviation can be used to capture this row-wise imbalance.

\textbf{Insight 3} further states how $N$ and the sparse matrix size affects the effectiveness of two principles. A benefit of parallel-reduction is to parallelize operations and data loading compared with sequantial version, but this becomes unnecessary when the workload is heavy and the resource occupancy is high. GPU has a limited amount of computation resources, and when the total workload is large, GPU performs works with multiple waves of threads. In this case, the workload imbalance is less serious a problem, since new threads will occupy the resource of the early-finishing threads.

\begin{figure}[t]
    \includegraphics[width=0.8\linewidth]{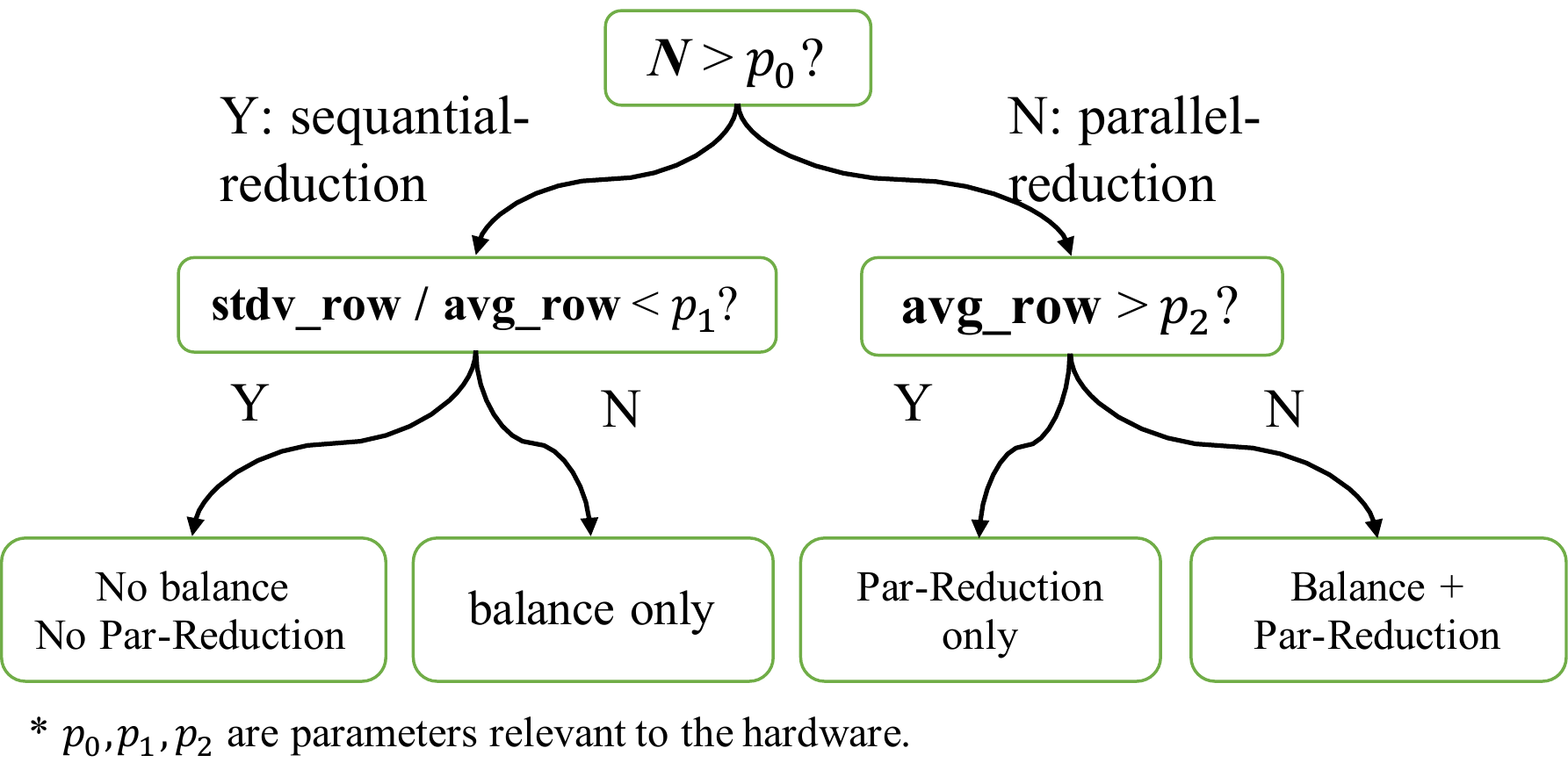}
    \caption{Adaptive strategy to select kernels.}
    \label{fig:decision}
\end{figure}

\begin{figure}[t]
    \includegraphics[width=\linewidth]{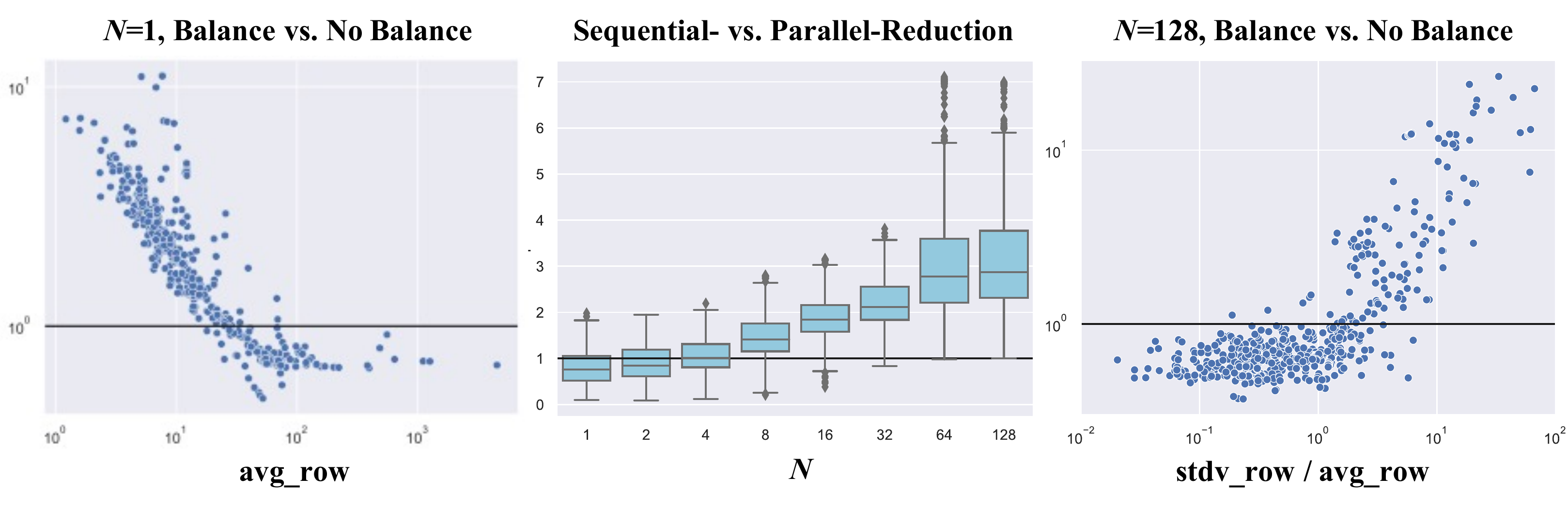}
    \caption{Validation of our adaptive strategy. 
    Middle: we observe parallel-reduction is beneficial only when $N$ is small. Left: When $N=1$, how workload-balancing benefit correlates with the average row-length grows. Right: When $N=128$, how workload-balancing benefit correlates with both the variance and average of row-lengths.
    }
    \label{fig:intuition}
\end{figure}

Our kernel selection strategy follows the three insights, and is shown in Figure.~\ref{fig:decision}. We make decisions with the following steps: First, we consider $N$ to choose between parallel- or sequential-reduction, following insight 1. We choose parallel-reduction for SpMV, and also SpMM with $N\leq 4$, when parallel-reduction can benefit from the vector-type data loading we propose in Section ~\ref{sec:approach:adaptive}. When $N>4$, we apply sequential reduction. Next, we decide whether to apply workload balancing according to sparse matrix features. According to insight 1, workload-balancing benefits matrices with imbalanced non-zero distribution in different rows. Hence, a high standard-deviation of the row-length, $stdv\_row$, is a positive signal for us to apply workload-balancing. According to insight 3, if the total amount of work is large, workload-imbalance becomes less serious. Thereby, a large mean row-length $avg\_row$ is a negative signal because it indicates a large number of non-zeros and amount of work. We combine the two signals and use $stdv\_row / avg\_row$ as the metric and empirically decide the threshold. Finally, for parallel-reduction kernels, we observe that a large $avg\_row$ greatly benefits the imbalanced parallel-reduction because short rows cause the idle discussed in our VSR optimization. Thereby, for parallel-reduction cases, we use $avg\_row$ to decide whether to apply workload-balancing. This completes our kernel selection strategy.

% We analyze the design options of SpMV/MM based on three insights.
% \begin{itemize}
%     \item 
%     \textbf{Insight 1:} The dataflow of SpMV/MM is composed by three nested loops, and the design space can be established through binding parallel workers to different loops.
%     \item  
%     \textbf{Insight 2:} The partition of M-loop involves a trade-off between workload-balance and work efficiency. 
%     \item 
%     \textbf{Insight 3:} The choice of the inner-most loop to parallelize, either N-loop or K-loop, involves a trade-off between efficient access of sparse-matrix and dense-matrix elements.
% \end{itemize}

% Approach: unified dataflow representation and scheduling options.
% Uniqueness: Abstract away sparse format and inner/outer product. Sparse unique. Compositing. 

% Approach: implementation of design points with optimizations.
% Design 1: parallel segmented reduction.
% Design 2: workload coarsening.
% Design 3: decoupled access and use.
% Uniqueness: first work to solve segmented reduction with warp-level primitives. First to reveal small N coarsening can be implemented by vector loading. First work to reveal scratchpad memory is better than warp shuffling. 

\section{Results} \label{sec:results}
\subsection{Kernel Performance}
\textbf{Platform}: We conduct experiments on three GPUs: Nvidia Tesla V100 (Volta architecture 7.0 compute capability), Nvidia RTX 2080 (Turing architecture 7.5 compute capability), Nvidia RTX 3090 (Ampere architecture 8.6 compute capability).

% \textbf{RTX 3090.} Nvidia RTX 3090, Compute Capability 8.6 (68 Ampere SMs at 1.395 GHz, 24 GB GDDR6X with 936 GB/s bandwidth). Host CPU: AMD Ryzen Threadripper 3970X (32 cores).

% \textbf{RTX 2080.} Nvidia RTX 2080, Compute Capability 7.5 (46 Turing SMs at 1.515 GHz, 8 GB GDDR6 with 448 GB/s bandwidth). Host CPU: Intel(R) Core(TM) i7-9700K (8 cores).

% \textbf{Tesla V100.} Nvidia Tesla V100, Compute Capability 7.0 (80 Volta SMs at 1.370 GHz, 16 GB HBM2 with 900 GB/s bandwidth). Host CPU: Intel(R) Xeon(R) CPU E5-2698 v4 (20 cores).

% \textbf{Benchmark and baselines} are detailed in each subsection.

We use SuiteSparse matrix collection~\cite{10.1145/2049662.2049663} to benchmark the kernel performance. We compare with two baselines: cuSPARSE in CUDA toolkit version 11.2, and ASpT~\cite{aspt}, the state-of-the-art SpMM implementation. We test $N$ from 1 up to 128. 
There are two ways to use our kernels: either to profile and select the best implementation off-line, or use our adaptive strategy for online selection. In most HPC and GNN applications, the sparse matrix can be profiled statically to select out the best kernel for iterative algorithms. We show the results of both approaches in Figure.~\ref{fig:kernel}. 

The ``ours'' in Figure.~\ref{fig:kernel} denotes the best of four implementations. For SpMV, our kernel is \textbf{1.14$\times$, 1.07$\times$, 1.11$\times$} against cuSPARSE on three GPUs respectively. For SpMM, our kernel exceeds cuSPARSE by  \textbf{1.26-1.41$\times$, 1.09-1.44$\times$ 1.22-1.57$\times$} on three GPUs respectively. Against ASpT~\cite{aspt} on settings they support, we achieve  \textbf{1.21$\times$ 1.14$\times$, 1.16$\times$} when $N=32$ and \textbf{1.18$\times$,1.14$\times$,1.06$\times$} when $N=128$.

\subsection{Adaptive Strategy}

\begin{figure*}[!tp]
    \includegraphics[width=0.9\linewidth]{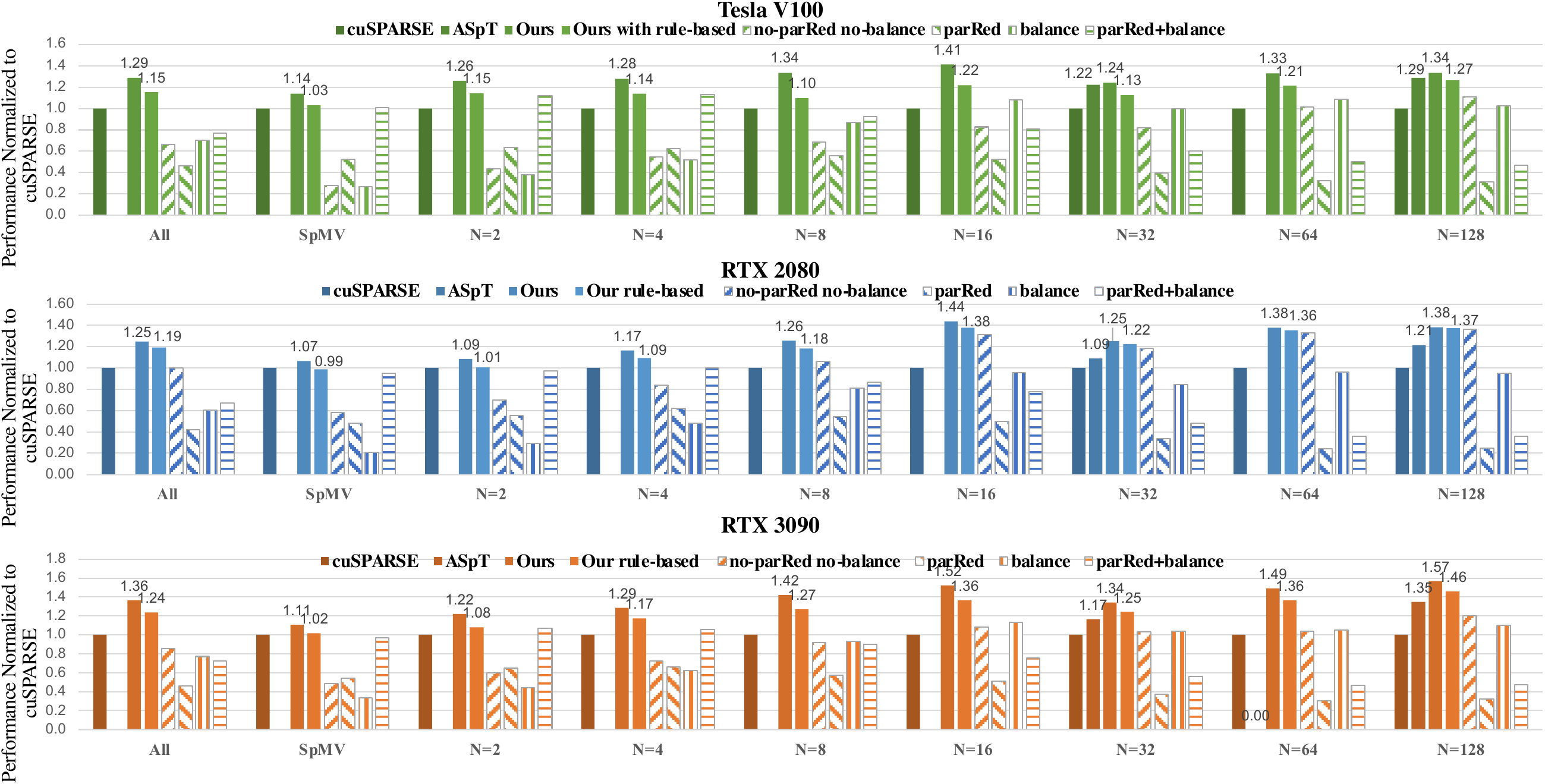}
    \caption{Kernel performance compared against cuSPARSE~\cite{cusparse} and ASpT~\cite{aspt}, tested on SuiteSparse benchmark~\cite{10.5555/829576}. }
    \label{fig:kernel}
\end{figure*}

The ``ours with rule-based'' in Figure.~\ref{fig:kernel} denotes the performance of kernels selected by our adaptive strategy, as stated in Section.~\ref{sec:approach:adaptive}. Compared with the optimal choice under different $N$, the kernel selected by our rules is \textbf{6\%-22\%, 1\%-8\%, 7\%-12\%} slower. Compared with the last four bars, which denote the four implementations, our strategy consistently outperforms the solution of always choosing the same kernel. In addition, we can observe that the best option among four individual kernels changes along with $N$. Thereby, when averaged among all $N$, the best single-kernel solution involves \textbf{68\%, 86\%, 76\%} performance loss, while our kernel selection rules involve only \textbf{12\%, 5\%, 10\%} performance loss.

\section{Related Work} \label{sec:rw}
 
\textbf{Categorized by optimization methods}: 

\underline{Workload-balancing:} The workload-balancing is achieved by row-binning, i.e. coarsely grouping sparse-matrix rows into buckets with similar total workload~\cite{greathouse2014efficient,sputnik}, or merge-path, i.e. fine-grained segmentation of arithmetic and memory operations to ensure balancing~\cite{merrill2016merge}. Yang \textit{et al.}~\cite{graphblas} extend MergePath to MergeSpmm, and in addition propose RowSplit which omits workload-balancing. The authors select between MergeSpmm and RowSplit according to an average length of sparse-matrix rows. We extend this approach to SpMM with arbitrary $N$, and improve the selection heuristics according to profiles on a more comprehensive benchmark.

\underline{Parallel-reduction:} Bell \& Garland~\cite{bell2009implementing} proposed the CSR-vector SpMV algorithm to load and compute on sparse elements in a vectorized fashion, in contrast to CSR-Scalar~\cite{bell2009implementing} which uses sequential reduction. CSR-Stream~\cite{greathouse2014efficient} exploits the parallel loading but sequential reduction of segments. Yang \textit{et al.}~\cite{graphblas} combines the vectorized loading with sequential reduction through a SIMD-shuffle primitive in CUDA. We greatly improved their implementation through exploiting the shared memory, as detailed in our prior publication~\cite{gespmm}. 

\underline{Specialized sparse format:} Compressed formats like ELL, block-CSR, HYB~\cite{cusparse} improves data access efficiency but at the cost of padded zeros and wasted computation. Specialized formats are also used to mark clustered elements and expose chances for data re-use~\cite{aspt}. Format-dedicated optimizations are orthogonal to the design space we explore in this paper. 

\textbf{Categorized by application}, the most-related prior work includes SpMM optimizations for graph neural networks~\cite{hu2020featgraph} and data analytics~\cite{hong2018efficient}. SpMV acceleration is also studied for scientific computing~\cite{aspt} and graph analytics~\cite{yang2011fast}. SpMV and SpMM, as alternatives to GEMV, GEMM in sparse DNN, is widely studies on GPU~\cite{gale2020sparse,wang2020sparsert}, CPU~\cite{elsen2020fast} and accelerators~\cite{eie,qin2020sigma}. Our work brings significant benefit to GNN and data analytics (matrix factorization), while also bring improvements the long-studied SpMV problem in scientific and graph domains.

\section{Conclusion} \label{sec:concl}

We extend the principles and practice of workload-balancing and parallel-reduction to a wide range of SpMV/MM problems. We propose three optimizations: VSR effectively combines the two techniques in SpMV, VDL and CSC optimize memory operation efficiency for SpMM with small and large $N$ respectively. Altogether, we provide highly optimized implementations of two design principles in SpMV/MM. The second main contribution is strategies to selectively apply the two techniques based on low-cost metrics such as row-length average and deviation. Optimal choices with our optimizations exceed the cuSPARSE library by \textbf{1.07-1.57$\times$} on different problems and GPUs. If the off-line profile is prohibited, our selection strategy involves only\textbf{ 5-12\% }performance loss on average, compared with a minimum 68\% if using a single kernel. We are collaborating with CogDL~\cite{cogdl}, a popular graph learning framework to apply our research.

\bibliographystyle{unsrt}
\bibliography{ref}

\end{document}